\documentclass[%
reprint,
superscriptaddress,
amsmath,amssymb,
aps,
pra,
]{revtex4-1}

\usepackage{graphicx}
\usepackage{dcolumn}
\usepackage{bm}
\usepackage{hyperref}

\usepackage{color}

\newcommand{\Ai}{\operatorname{Ai}}
\newcommand{\sinc}{\operatorname{sinc}}
\newcommand{\sgn}{\operatorname{sgn}}
\newcommand{\dif}{\operatorname{d}}
\newcommand{\sig}{\operatorname{sig}}

\begin{document}

\title{Precise amplitude, trajectory, and beam-width control of accelerating and abruptly autofocusing beams}

\author{Michael Goutsoulas}
\affiliation{Department of Mathematics and Applied Mathematics, University of Crete, 70013 Heraklion, Crete, Greece}
\author{Nikolaos K. Efremidis}%
\email{nefrem@uoc.gr}
\homepage{http://www.tem.uoc.gr/~nefrem}
\affiliation{Department of Mathematics and Applied Mathematics, University of Crete, 70013 Heraklion, Crete, Greece}
\affiliation{Institute of Applied and Computational Mathematics, Foundation for Research and Technology - Hellas (FORTH), 70013 Heraklion, Crete, Greece.}





\date{\today}

\begin{abstract}
We show that it is possible to independently control both the trajectory and the maximum amplitude along the trajectory of a paraxial accelerating beam. This is accomplished by carefully engineering both the amplitude and the phase of the beam on the input plane. Furthermore, we show that the width of an accelerating beam is related only on the curvature of the trajectory. Therefore, we are able to produce beams with predefined beam widths and amplitudes. These results are useful in applications where precise beam control is important. 
In addition we consider radially symmetric abruptly autofocusing beams. We identify the important parameters that affect the focal characteristics. Consequently, we can design autofocusing beams with optimized parameters (such as sharper focus and higher intensity contrast). In all our calculations the resulting formulas are presented in an elegant and practical form in direct connection with the geometric properties of the trajectory. 
Finally we discuss methods that can be utilized to experimentally realize such optical waves. 
\end{abstract}

\maketitle


\section{Introduction}

Over the last decade the study of optical beams with engineered trajectories has been very successful in generating novel classes of waves for particular applications. The research in this field initiated with the prediction and experimental observation of accelerating diffraction-free Airy beams~\cite{sivil-ol2007,sivil-prl2007}. By engineering the phase profile of the optical wave it is shown that paraxial classes of curved beams with predefined arbitrary convex trajectories can be generated~\cite{green-prl2011,chrem-ol2011,froeh-oe2011}. Using a different approach it is possible to generate Bessel-like beams that can even bend along non-convex type of trajectories~\cite{chrem-ol2012-bessel,zhao-ol2013}. Accelerating waves in the non-paraxial regime have a main advantage that the trajectory of the beam can bend at large angles~\cite{froeh-oe2011,kamin-prl2012,courv-ol2012,zhang-ol2012,zhang-prl2012,aleah-prl2012,bandr-njp2013,penci-ol2015}. The curved trajectory and self-healing characteristics of such optical waves have been proven very useful in a variety of applications ranging from filamentation~\cite{polyn-science2009,polyn-prl2009} and electric discharge generation~\cite{cleri-2015} to particle manipulation~\cite{baumg-np2008,zhang-ol2011,zheng-ao2011,schle-nc2014,zhao-sr2015}, microscopy and imaging~\cite{jia-np2014,vette-nm2014}, and micromachining~\cite{jia-np2014,vette-nm2014}. Accelerating waves have been utilized in generating an abrupt wave focusing or abrupt autofocusing by an on-axis collapse of a ring-shaped caustic~\cite{efrem-ol2010}. The maximum intensity of such beams remains almost constant up until the focus where it abruptly increases by orders of magnitude. Abruptly autofocusing waves have been utilized in particle manipulation~\cite{zhang-ol2011}, creating ablation spots in materials~\cite{papaz-ol2011}, and filamentation~\cite{panag-nc2013}. In the non-paraxial regime abruptly autofocusing waves are associated with increased intensity contrast~\cite{penci-ol2016}. 

In the bibliography  there are some works that discuss particular cases of amplitude manipulation of accelerating beams~\cite{hu-ol2013-am,penci-ol2015}. However, most of the effort up to this point has been devoted in engineering the trajectory and does not take into account other important beam parameters, such as the amplitude and the beam width. In particular, no systematic method has been developed for engineering these two very important beam parameters. Note that, for example, in particle manipulation it is important that the curved beam maintains a constant maximum intensity so that the particles get transported without interruptions. 

The purpose of this work is to generate beams with judiciously designed properties (trajectory, amplitude, and width) in the paraxial domain. This is accomplished by engineering both the amplitude and the phase of the beam on the input plane. Specifically, we show that the beam width is solely related with the curvature of the trajectory. In addition, the maximum amplitude along the trajectory is related with both the geometric properties of the trajectory and the amplitude of the beam on the input plane. As a result, accelerating beams can have arbitrary predefined convex trajectories (and thus designed beam widths) and engineered maximum amplitude. The only requirement is that the amplitude on the input plane is relatively slowly varying. We also analyze the focusing characteristics of abruptly autofocusing beams. We find analytic expressions for the trajectory and the maximum amplitude along the trajectory which can be utilized to engineer autofocusing beams with optimal characteristics (sharper focus, maximum contrast). The resulting mathematical formulas are expressed in an elegant and practical form in connection to the geometric properties of the beam trajectory. Finally, we discuss about methods that can be utilized to generate such beams with designed characteristics.

\section{Amplitude-trajectory engineering of accelerating beams\label{sec:1}}

The dynamics of an optical beam propagating in one transverse dimension is governed by the Fresnel diffraction integral 
\begin{equation}
\psi(x,z)=\frac{1}{(i \lambda z)^{1/2}}\int_{-\infty}^{\infty}\psi_0(\xi)\exp\left [ik \frac{(x-\xi)^2}{2z} \right ]\,\dif\xi,
\label{eq:Fresnel}
\end{equation}
where $x$ is the transverse and $z$ the longitudinal propagation direction, $k=2\pi n\nu/c=2\pi/\lambda$, $c$ is the speed of light, $n$ is the refractive index, $\nu$ is the optical frequency, $\lambda$ is the wavelength in the dielectric medium, and $\psi_0(x)$ is the optical wave excitation on the input plane ($z=0$). By decomposing $\psi_0$ into amplitude and phase as $\psi_0(x)=A(x)e^{i\phi(x)}$ we obtain the total phase $\Psi$ that is involved in the Fresnel integrand 
\begin{equation}
\Psi(\xi;x,z) = \phi(\xi) + k\frac{(x-\xi)^2}{2z}. 
\label{eq:Psi}
\end{equation}
In terms of catastrophe theory~\cite{kravt-1999,berry-1980}, $\Psi$ is the potential, $x$, $z$ are the control variables, $\xi$ is the internal variable, and $\partial_\xi\Psi=0$ is the surface of equilibria. The catastrophe condition in the case of one internal variable consists of the points that lie in the surface of equilibria and satisfy $\partial_{\xi\xi}\Psi=0$. Following the relevant calculations from the surface of equilibria we derive the ray equation $x=\xi+\phi'(\xi)z/k$. In addition, from the catastrophe condition we obtain the high intensity beam trajectory
\begin{equation}
(x_c(\xi_c),z_c(\xi_c))=
\left(\xi_c-\frac{\phi'(\xi_c)}{\phi''(\xi_c)},-\frac k{\phi''(\xi_c)}\right)
\label{eq:causticphase}
\end{equation}
as a function of the phase on the input plane. Note that the subscript $c$ in the formulas stands for caustic. Importantly, we can also solve the inverse problem of determining the required phase as a function of the convex by otherwise arbitrary predefined trajectory
\begin{equation}
x_c=f(z_c). 
\label{eq:trajectory}
\end{equation}
In particular, we have to take into account that the line that is tangent at each point of the trajectory is described by a ray equation and thus
\begin{equation}
\frac{\dif\phi}{\dif\xi} = k \frac{df(z_c(\xi))}{dz_c}
\label{eq:tangent}
\end{equation}
where $z_c(\xi)$ is obtained from
\begin{equation}
\xi=f(z_c)-z_cf'(z_c). 
\label{eq:xiofzc}
\end{equation}
Note that since $z_c=-k/\phi''(\xi_c)>0$ ($z=0$ is the incident plane), a caustic is formed only when $\phi''(\xi)<0$. 

In order to obtain an expression for the amplitude close to the caustic we perturb the variables $x$ and $\xi$ with respect to their values at the caustic~\cite{chrem-ieee2013}
\begin{equation*}
x=x_c+\delta x, \quad \xi=\xi_c+\delta\xi
\end{equation*}
while we keep a constant value for $z=z_c$. We then expand the phase $\Psi(x_c+\delta x,z_c,\xi_c+\delta\xi)$ is a Taylor series and keep all the terms up to cubic order [i.e., $(\delta x)^j(\delta\xi)^k$, $j+k\le3$]. Importantly, 
we assume that the amplitude $A$ is not constant but is slowly varying with $\xi$. In our calculations, due to phase stationarity at $\xi=\xi_c$ we assume that $A(\xi_c+\delta\xi)\approx A(\xi_c)$. Integrating with respect to $\delta\xi$ leads to
\begin{equation}
\psi = 
2A(\xi)
\left(
\frac{\pi^4 z_c^3\kappa^2}{\lambda}
\right)^{1/6}
e^{i\Xi}
\Ai(s(2k^2\kappa)^{1/3}\delta x),
\label{eq:psixz}
\end{equation}
where
\begin{equation}
\Xi = 
\phi
+\frac{k(x_c-\xi)^2}{2z_c}
+\frac{k(x_c-\xi)}{z_c}\delta x +\frac k{2z_c}(\delta x)^2-\frac\pi4,
\label{eq:Xi}
\end{equation}
\[
\kappa(z_c)=\left|\frac{\dif^2f(z_c)}{\dif z_c^2}\right|
\]
is the curvature of the trajectory in the paraxial approximation, $s=\sgn(\dif^2f(z_c)/\dif z_c^2)$ is the sign of the curvature, $\Ai$ is the Airy function, and for simplicity we have replaced $\xi_c$ with $\xi$.

Let us utilize Eq.~(\ref{eq:psixz}) to obtain some significant information about the properties of the beam close to the caustic. 
First of all we note that independently of the functional form of the selected trajectory, close to the caustic the optical wave is described by an Airy function that varies linearly with $\delta x$. It is interesting to point out that the beam width 
\[
w(z) = \frac1{(2k^2\kappa(z))^{1/3}}
\]
depends solely on the paraxial beam curvature of the trajectory. Specifically, the beam width is inversely proportional to the cubic root of the curvature. We conclude that a beam has constant width if and only if the curvature of the trajectory is constant. Thus the only class of accelerating waves with constant width are those of the Airy-type that follow a parabolic trajectory of the form $x_c=c_0+c_1z_c+c_2z_c^2$ with $c_j$ being arbitrary constants. We can generalize the above statement by saying that a beam has constant width as long as its trajectory remains parabolic.
The other important parameter is the beam amplitude. From Eq.~(\ref{eq:psixz}) we note that it depends linearly on the amplitude on the incident plane $A(\xi)$. In addition, it depends on the geometric properties of the trajectory: it increases as we increase the beam curvature $\kappa$ and the distance from the incident plane $z_c$. 

\begin{figure*}[t]
\includegraphics[width=\textwidth]{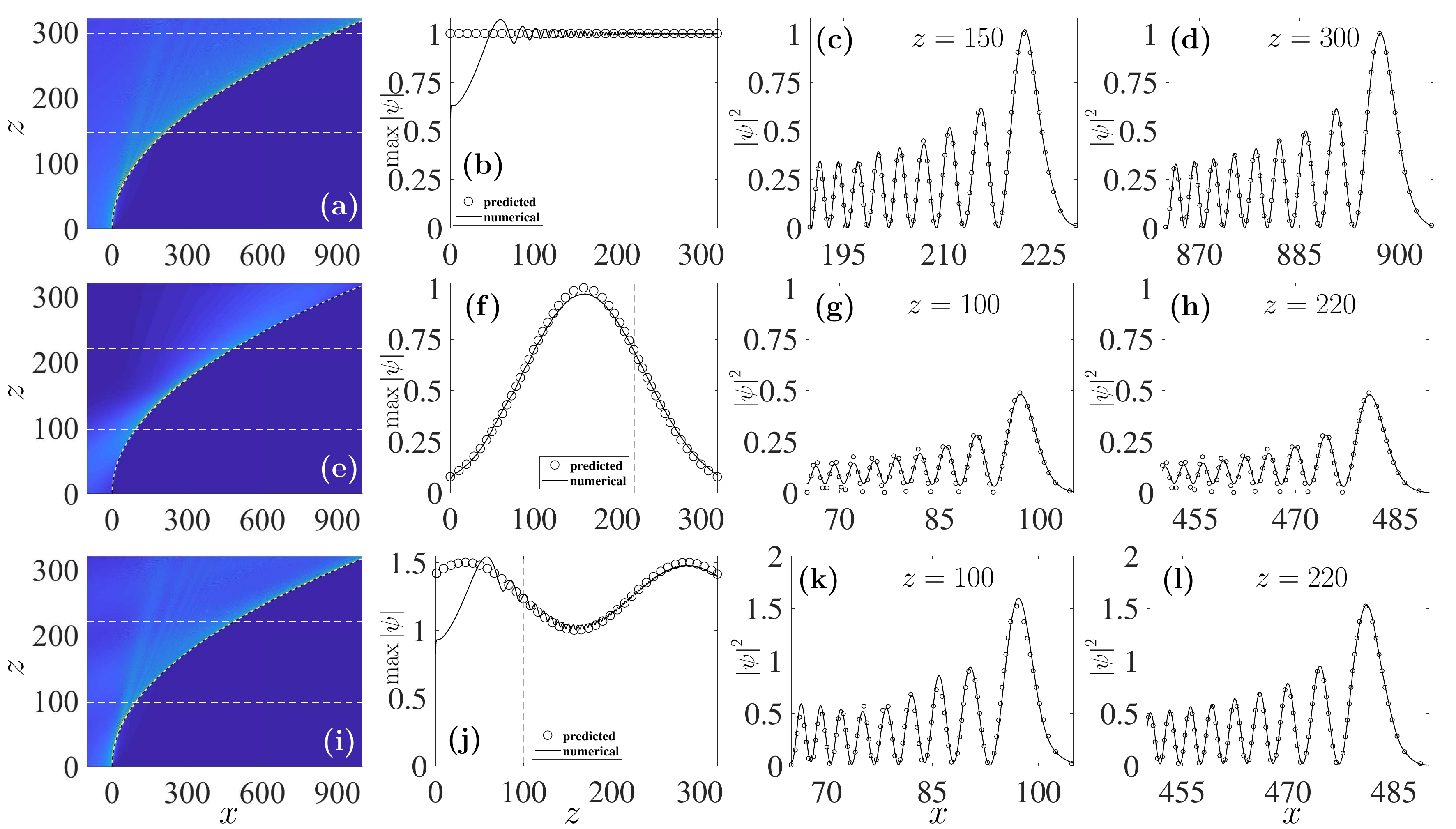} 
\caption{Accelerating beams following a parabolic trajectory [Eq.~(\ref{eq:powerlaw}) with $\alpha=2$ and $\beta=0.01$]. In the first column we can see the amplitude dynamics and the theoretical prediction for the trajectory (dashed curve). In the second column the maximum amplitude as a function of the propagation distance (solid curve) along with the theoretical prediction (shown in circles) is depicted. 
In the third and forth columns cross sections of the beam intensity at different propagation distances are presented with the theoretical predictions shown in circles. 
In the three rows the theoretical maximum value of the field amplitude is $U(z)=1$,  $U(z)=\exp(-(z-160)^2/100^2)$, and $U(z)=1+0.5\sin^2((z-160)/80)$, respectively. The horizontal dashed lines in the first column correspond to the cross sections shown in the third and forth columns.}
\label{fig:1}
\end{figure*}
We can utilize the phase on the input plane to design beams with predefined trajectory or beam-width. The amplitude on the input plane $A(\xi)$ provides an additional degree of freedom that can be employed to engineer the maximum beam amplitude 
\[
U(z) = \{\max(|\psi(x,z)|):x\in\mathbb{R}\}.
\]
Specifically from Eq.~(\ref{eq:psixz}) we obtain an explicit relation for the required amplitude on the input plane 
\begin{equation}
A(\xi) = 
\frac{U(z_c(\xi))}{2.3}
\left(
\frac{\lambda}{\kappa^2z_c^3}
\right)^{1/6}. 
\label{eq:ampl}
\end{equation}
Direct numerical simulations presented below confirm the accuracy of this formula. We conclude that both the trajectory/beam width and the maximum amplitude along the trajectory can be pre-engineered provided that we utilize both the amplitude and the phase of the beam on the input plane. 

Although analytic expressions for different classes of convex trajectories can be found in closed form, for the purposes of this study we restrict ourselves to the case of power law trajectories~\cite{green-prl2011,chrem-ol2011,froeh-oe2011}
\begin{equation}
x_c=f(z_c)=\beta z_c^\alpha
\label{eq:powerlaw}
\end{equation}
with $\alpha>1$. Following the relevant calculations we obtain the required phase profile on the input plane 
\begin{equation}
\phi(\xi) = 
\frac{-k\beta^{1/\alpha}\alpha^2}{(\alpha-1)^{1-1/\alpha}}
\frac{(-\xi)^{2-1/\alpha}}{2\alpha-1}
\label{eq:powerlawphase}
\end{equation}
where $\xi<0$. For an arbitrary but relatively slowly varying amplitude profile $U(z_c)$ along the caustic we derive the following prediction
\begin{equation}
A(z_c(\xi)) = 
\frac{U(z_c)}{2.3}
\left(
\frac{\lambda} { [\alpha(\alpha-1)\beta]^2 z_c^{(2\alpha-1)}}
\right)^{1/6}
\label{eq:A}
\end{equation}
for the required amplitude on the input plane, where $z_c(\xi)=[-\xi/(\beta(\alpha-1))]^{1/\alpha}$. A particularly interesting case is that of constant amplitude along the trajectory $U(z_c) = c$. From Eq.~(\ref{eq:A}) we see that this is possible by selecting $A(\xi)\propto1/(-\xi)^{(2\alpha-1)/(6\alpha)}$. In the particular case of a parabolic trajectory we recover the characteristic amplitude profile $A(\xi)\propto 1/(-\xi)^{1/4}$ of the Airy function. Finally, the width of the beam is given by
\begin{equation*}
w(z)=
\frac1{[2k^2\beta\alpha(\alpha-1)z^{\alpha-2}]^{1/3}}.
\end{equation*}
We see that for $\alpha=2$ the beam width remains (as expected) invariant, for $\alpha>2$ the beam width decreases with $z$, whereas for $1<\alpha<2$ the beam width increases with $z$.

\begin{figure*}
\includegraphics[width=\textwidth]{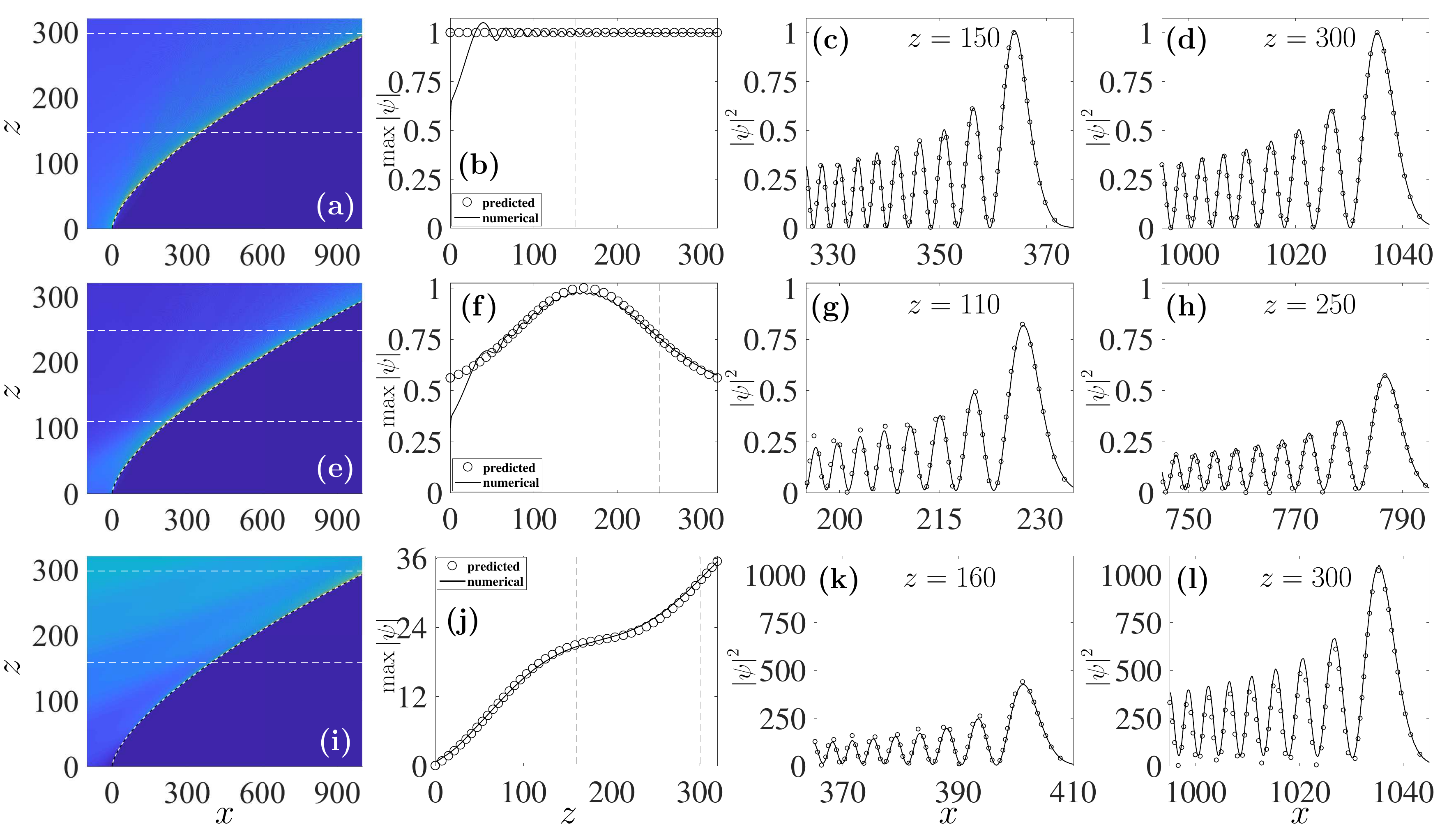} 
\caption{Accelerating beams following a power law trajectory [Eq.~(\ref{eq:powerlaw}) with $\alpha=3/2$ and $\beta=1/5$]. In the first column we can see the amplitude dynamics and the theoretical prediction for the trajectory (dashed curve). In the second column the maximum intensity as a function of the propagation distance (solid curve) along with the theoretical prediction (shown in circles) is depicted. 
In the third and forth columns cross sections of beam intensity at different propagation distances are presented with the theoretical predictions shown in circles. 
In the three rows the theoretical maximum value of the field amplitude is $U(z)=1$, 
$U(z)=0.5+0.5\exp(-(z-160)^2/110^2)$, and $U(z)=0.1z+6\sin^2(z/80)$, respectively. The horizontal dashed lines in the first column correspond to the cross sections shown in the third and forth columns.}
\label{fig:2}
\end{figure*}
In our simulations we use normalized coordinates. Specifically, we scale the transverse coordinates with respect to $x_0$ (i.e., $x\rightarrow x_0x$), the longitudinal variables with respect to $kx_0^2$ (i.e. $z\rightarrow kx_0^2z$), and the amplitude to an arbitrary scaling. With these substitutions, all the parameters in the formulas derived in this section become normalized and dimensionless with the simple replacement $k\rightarrow 1$. In Fig.~\ref{fig:1} we present results in the case of a parabolic trajectory [Eq.~(\ref{eq:powerlaw}) with $\alpha=2$ and $\beta=10^{-2}$]. In the three rows different functional forms for the maximum amplitude along the caustic $U(z_c)$ are selected. In particular, in the first row the maximum amplitude is constant, in the second row a Gaussian profile is selected, while in the third row the amplitude is the sum of a constant and a sinusoidal function. We note that in all cases the theoretical prediction $U(z)$ is in excellent agreement with the numerical results. However, in the first and the third row we see a transition distance before the maximum amplitude reaches the theoretical value. 
Specifically, at the first stages of propagation the numerical value of the amplitude is smaller than expected but gradually increases and reaches the theoretical engineered profile.  
This distance is negligible in the second row where the maximum amplitude exhibits smooth changes due to its Gaussian profile. In the last two columns of Fig.~\ref{fig:1} we compare the numerically computed amplitude profile at different cross sections with the theoretical prediction given by Eq.~(\ref{eq:psixz}). We see that the agreement is excellent not only in describing the main lobe but also for several additional lobes both as it concerns the frequency and the amplitude of the oscillations. This result is surprisingly accurate taken into account that in our calculations $\delta x$ is taken to be small. 

\begin{figure*}
\includegraphics[width=\textwidth]{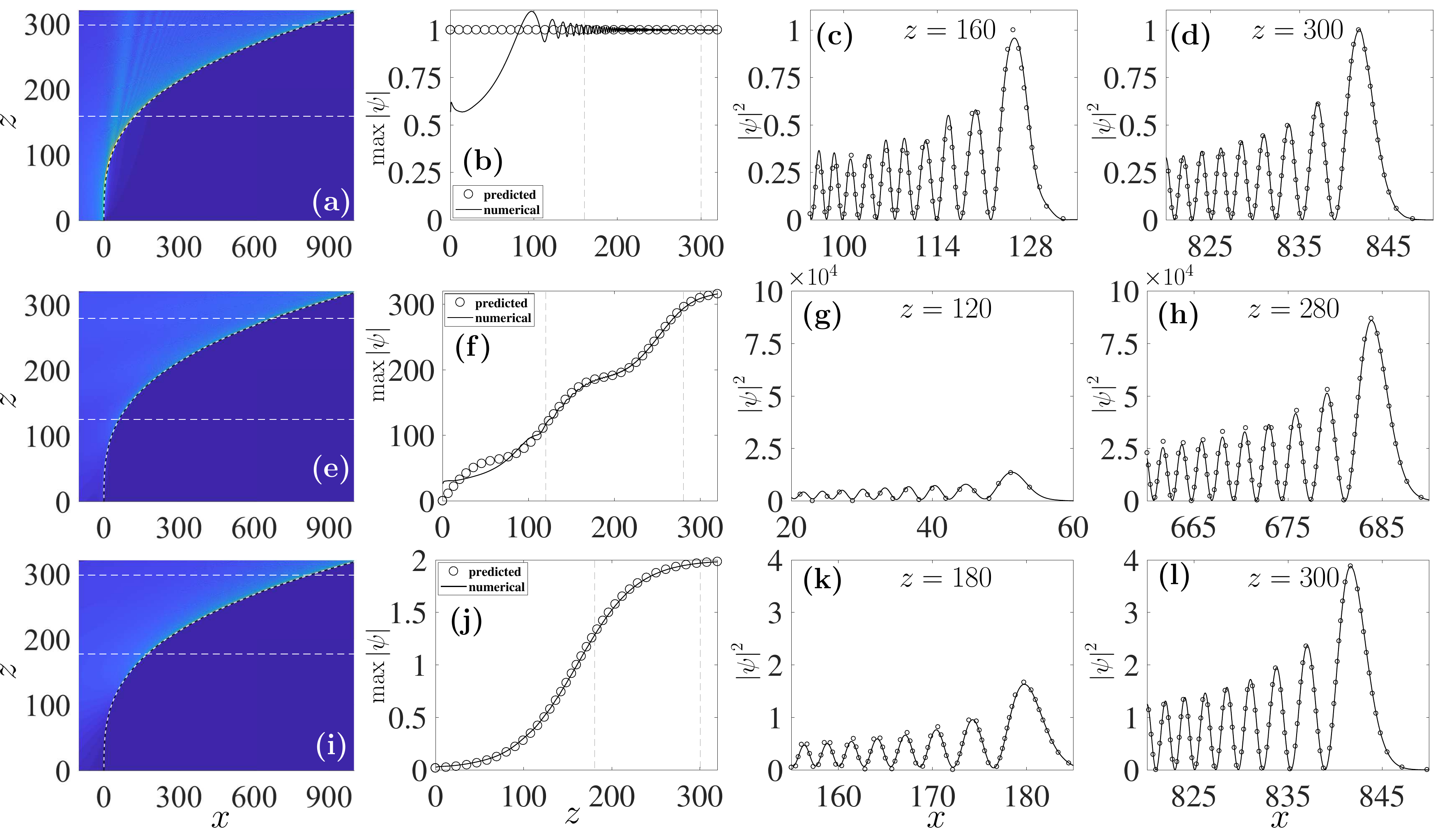} 
\caption{Accelerating beams following a cubic trajectory [Eq.~(\ref{eq:powerlaw}) with $\alpha=3$ and $\beta=1/32000$]. In the first column we can see the amplitude dynamics and the theoretical prediction for the trajectory (dashed curve). In the second column the maximum intensity as a function of the propagation distance (solid curve) along with the theoretical prediction (shown in circles) is depicted. 
In the third and forth rows cross sections of beam intensity at different propagation distances are presented with the theoretical predictions shown in circles. 
In the three rows the predicted maximum value of the field amplitude is $U(z)=1$, $U(z)=z+ 15\sin(z/20)$, and $U(z)=1+\tanh(.015(z-160/2))$, respectively.
The horizontal dashed lines in the first column correspond to the cross sections shown in the third and forth columns.}
\label{fig:3}
\end{figure*}
To highlight the potential of our method we also obtain results in the case of power law trajectories with different exponents and a selection of different amplitude profiles $U(z)$. Specifically, in Fig.~\ref{fig:2} the exponent is $\alpha=3/2$ and $U(z)$ is constant in the first row, an elevated Gaussian in the second row, and sinusoidal with an additional linear term in the third row. Finally, in Fig~\ref{fig:3} we select a cubic trajectory. In the three rows the amplitude along the trajectory is constant, sinusoidal with an additional linear term, and a sigmoid function, respectively. Comparing these two cases, we can see that the transition region for the amplitude in Fig.~\ref{fig:2} is smaller as compared to Fig.~\ref{fig:3}. 
For example, in the first row of these figures the expected amplitude profile is constant and unitary. The numerically computed amplitude converges to the theoretical at $z=60$ in Fig.~\ref{fig:2} and at $z=100$ in Fig.~\ref{fig:3}. This happens because during the early stages of propagation the paraxial curvature $\kappa$ is smaller along the trajectory of Fig.~\ref{fig:3} as compared to Fig.~\ref{fig:2}.

\section{Amplitude-trajectory engineering of abruptly autofocusing beams\label{sec:2}}

In the case of abruptly autofocusing waves the radial symmetry of the beam results in the following Fresnel-type diffraction integral
\begin{equation}
\psi(r,\theta)=
\frac{k e^{i\frac{kr^2}{2z}}}{iz}
\int_0^\infty \rho \psi_0(\rho)J_0\left(\frac{kr\rho}{z}\right)
e^{i\frac{k\rho^2}{2z}}\,\dif\rho
\label{eq:rad:Fresnel}
\end{equation}
where $r$, $\rho$ are radial coordinates 
and $\psi_0(r)=A(r)e^{i\phi(r)}$ is the field profile on the incident plane and its amplitude and phase decomposition. Using large argument asymptotics for the Bessel function
\begin{equation}
J_0(x) \approx \sqrt{\frac{1}{2i\pi x}}\left(
e^{ix}+ie^{-ix}
\right)
\label{eq:J0:asymptotic}
\end{equation}
and utilizing first and second order stationarity of the phase we derive the equations for the rays and the caustics. From these equations we can solve both the direct and inverse problem between the phase on the incident plane and the trajectory of the beam. The resulting equations are identical to Eqs.~(\ref{eq:causticphase})-(\ref{eq:xiofzc}) with the substitutions $x\rightarrow r$ and $\xi\rightarrow\rho$. There is a clear physical picture behind this equivalence: The rays propagate in a linear fashion and can not distinguish between Cartesian and radial coordinates.

We would like to utilize Eq.~(\ref{eq:rad:Fresnel}) in order to derive analytic expressions for the amplitude of the AAF beam close to the caustic. We follow a similar approach as in Section~\ref{sec:1} by using the expansion 
$r=r_c+\delta r$, $\rho=\rho_c+\delta\rho$, $z=z_c$ close to the caustic and large argument asymtptics 
for the Bessel function [Eq.~(\ref{eq:J0:asymptotic})].
We keep all the terms in the phase of the integrand in Eq.~(\ref{eq:rad:Fresnel}) up to cubic order [$(\delta\rho)^j(\delta r)^k$ with $j+k\le3$]. The resulting expression reads 
\begin{equation}
\psi = 
2A(\rho)
\sqrt{\frac{\rho}{ir}}
\left(
\frac{\pi^4 z_c^3\kappa^2}{\lambda}
\right)^{1/6}
e^{i\Xi}
\Ai(-(2k^2\kappa)^{1/3}\delta r)
\label{eq:aaf:drhodr}
\end{equation}
where
\[
\Xi = 
\phi(\rho)
+\frac{k(r_c-\rho)^2}{2z_c}
+kg(z_c)\delta r,
\]
\begin{equation}
r_c = \rho+\frac{\phi'(\rho)}{k}z_c,\quad
z_c = -\frac{k}{\phi''(\rho)},
\label{eq:aaf:asy:002:caustic}
\end{equation}
$g=\dif f_c(z_c)/\dif z_c$ is the slope of the trajectory. In the above formulas we have taken the second derivative of the trajectory to be negative ($s=-1$).
For simplicity, in the equations above we replaced $\rho_c\rightarrow\rho$. Note that we are going to use the same replacement in the formulas derived in the rest of this section. 
As in the case of accelerating waves the resulting expression depends on the geometric properties of the trajectory as well as on the amplitude on the input plane. 
Due to our assumption that the argument of the Bessel function is relatively large ($kr\rho/z\gg1$) the above equation diverges as $r\rightarrow0$ and thus fails to describe the optical wave close to the focus. 
However, Eq.~(\ref{eq:aaf:drhodr}) is very useful in describing the amplitude profile in the transverse plane before the wave focuses ($0<z<z_f$).

An expansion that works both at the early stages of propagation as well as close to the focus is $r=r_c$, $z=z_c+\delta z$, $\rho=\rho_c+\delta\rho$. This is a two-stage process followed by a global asymptotic expression. At a first stage, using the same methodology as before and assuming that $kr\rho/z\gg1$ we obtain
\begin{multline}
\psi(r)=
A(\rho)
\left(
\frac{2\kappa}{k}\right)^{1/3}
\left(\frac{2\pi k\rho z_c^2}{ir_cz}\right)^{1/2}e^{i\Xi}\\
\Ai\left(
(2k^2\kappa)^{1/3}g(z_c)\delta z
\right),
\label{eq:aaf:asy:002}
\end{multline}
where
\begin{equation}
\Xi=
\phi(\rho)+\frac{k(\rho-r_c)^2}{2z_c}
-\frac{kg^2(z_c)}{2}\delta z.
\label{eq:aaf:asy:002:phase}
\end{equation}
Due to our assumption that $kr\rho/z\gg1$ this formula also becomes inaccurate close to the optical axis: The denominator in Eq.~(\ref{eq:aaf:asy:002}) is proportional to $\sqrt{r}$ and thus the amplitude diverges as $r\rightarrow0$. Since the caustic approaches the axis when $z$ approaches the focal distance $z_f$ we conclude that Eq.~(\ref{eq:aaf:asy:002}) is valid for $0<z<z_f$. 

At a second stage, we use the same expansion $\rho=\rho_c+\delta\rho$, $z=z_c+\delta z$ but now with $r=0$, in order to derive an expression that is valid close to the focus. We employ a similar methodology as in the previous cases. Keeping the dominant terms we end up with  
\begin{equation}
\psi(0,\delta z) = 
A
\left(\frac{2\kappa}k\right)^{1/3}
\frac{2\pi k\rho z_c}{iz}e^{i\Xi}
\Ai
\left(
(2k^2\kappa)^{1/3}g\delta z
\right)
\label{eq:aaf:asy:005}
\end{equation}
where Eqs.~(\ref{eq:aaf:asy:002:caustic}), (\ref{eq:aaf:asy:002:phase}) are still valid (with $r_c=0$).

By inspection of the two asymptotic expressions given by Eqs.~(\ref{eq:aaf:asy:002}) and (\ref{eq:aaf:asy:005}) we see that they have the same argument inside the Airy function and the same phase factor. Thus, their only difference lies in the amplitude that multiplies the Airy function. There are several ways to combine these two formulas into a single global asymptotic expression. We select the following 
\begin{equation}
\psi =
A(\rho)e^{i\Xi} 
\frac{(2\kappa/k)^{1/3} 2\pi kz_c\rho} {\left[2\pi i k \rho zr_c-z^2\right]^{1/2}}
\Ai\left(
(2k^2\kappa)^{1/3}g\delta z
\right)
\label{eq:aaf:global}
\end{equation}
for its simplicity and accuracy. Equation~(\ref{eq:aaf:global}) is valid close to the caustic and for propagation distances $z$ that can even exceed by a small amount the focal distance $z_f$. Depending on the magnitude of $2\pi k\rho r_c/z_c$ the asymptotic expressions of Eqs.~(\ref{eq:aaf:asy:002}), (\ref{eq:aaf:asy:005}) are recovered. Assuming that the terms that contribute to the amplitude in Eq.~(\ref{eq:aaf:global}) are slowly varying functions of $\rho$ in comparison to the Airy function, we can estimate the location $z_f$ and the intensity of the focus by setting $r_c=0$ and the argument of the Airy function to $-1$. We see that the focal distance
\begin{equation}
z_f = z_c -\frac{1}{(2k^2\kappa(z_c))^{1/3}g(z_c)}
\label{eq:zf}
\end{equation}
is shifted from $z_c$ by an amount that is inversely proportional to the slope and the curvature of the trajectory (note that $g(z_c)<0$). 
The maximum field amplitude at the focus is then given by the following estimate 
\begin{equation}
|\psi_{\mathrm{max}}(z_f)|\approx2\pi \rho  A(\rho) (2k^2\kappa)^{1/3}\Ai(-1). 
\label{eq:foc_ampl}
\end{equation}
We conclude that there are only three fundamental parameters that affect the intensity of the beam at the focus: The amplitude $A(\rho)$ and the distance from the axis on the input plane $\rho$ of the ray that converges to the focus, and the curvature of the beam at the focus. 

We still need to compute the amplitude profile after the focus $z>z_f$. Interestingly, in this regime, the maximum amplitude does not lie close to the caustic as in the previous cases: At the focus a beam transformation takes place with a consequence that the maximum amplitude of the optical wave lies in a region close to the optical axis. For our calculations it is sufficient to assume that $kr\rho/z$ is relatively small for $z>z_f$. Applying first order stationarity of the phase in Eq.~(\ref{eq:rad:Fresnel}) we obtain
\[
\frac{k\rho}{z}+\phi'(\rho)=0.
\]
The above equation supports two solutions provided that they are smaller than the aperture $r_a$ (i.e. $\rho_1<\rho_2<\rho_a$). Defining by $z_{c,j}=-k/\phi''(\rho_{j})$ the location in the longitudinal direction where the rays emitted from $\rho_j$ contribute to the caustic we have that $z_{c,1}<z<z_{c,2}$. Thus one of the rays contributes to the caustic before and the other after the selected value of $z$. Using a stationary phase method we obtain
\begin{multline}
\psi(r,z)=
\sum_{j=1,2}
\rho_jA(\rho_j)
\left|\frac{2\pi kz_{c,j}} {z(z_{c,j}-z)}\right|^{1/2}
J_0\left(\frac{kr\rho_j}{z}\right) 
\\
e^{i[\frac{k(r^2+\rho_j^2)}{2z}+\phi(\rho_{j})+(\mu_j-2)\frac\pi4]}
\label{eq:aaf:asy:006}
\end{multline}
where
\[
\mu_j = \sgn(z_{c,j}-z),
\]
and thus $\mu_1=-1$, $\mu_2=1$. The above equation holds for $z\ge z_f$ and for relatively small values of $kr\rho/z$. An analytic expression for the wave amplitude can be obtained by defining the amplitude $C_j$ and the phase $\theta_j$ of the two terms in Eq.~(\ref{eq:aaf:asy:006}) and utilizing the formula $|C_1e^{i\Theta_1}+C_2e^{i\Theta_2}|=(C_1^2+C_2^2+2C_1C_2\cos(\Theta_1-\Theta_2))^{1/2}$. Due to destructive interference in Eq.~(\ref{eq:aaf:asy:006}) the maximum intensity is not always located exactly at the origin. However, it can always be found in an area that is close to the optical axis. After $\rho_2$ exceeds the aperture $\rho_2>\rho_a$ only the first term ($j=1$) is involved in Eq.~(\ref{eq:aaf:asy:006}) and from this point on the maximum amplitude is always located exactly at the origin
\[
|\psi_{\mathrm{max}}(z)| = 
\rho_1A(\rho_1)
\left|\frac{2\pi kz_{c,1}} {z(z_{c,1}-z)}\right|^{1/2}.
\]

\begin{figure}[!ht]
\includegraphics[width=0.9\columnwidth]{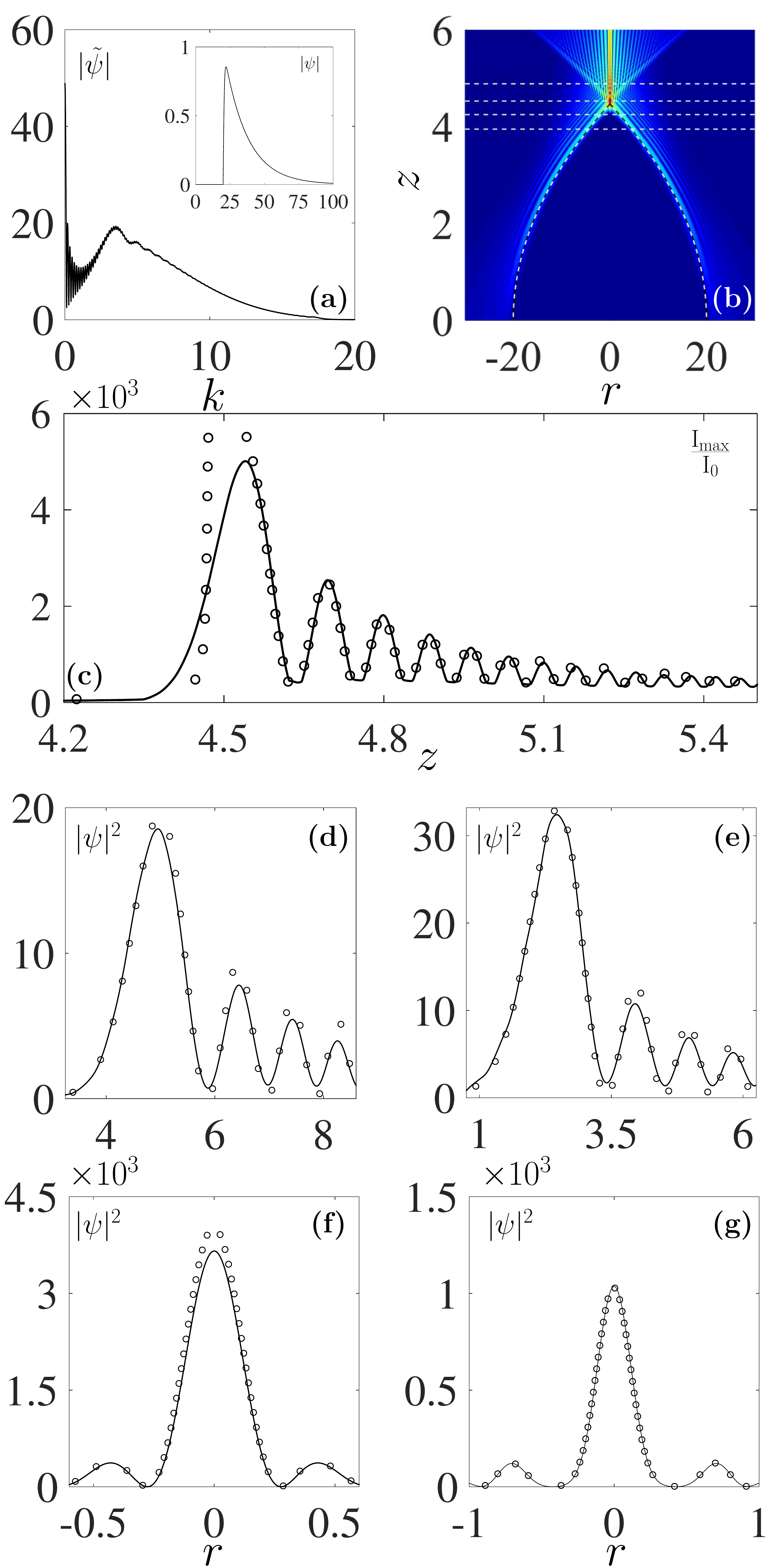} 
\caption{An abruptly autofocusing beam following a parabolic trajectory ($\alpha=2$) with $\beta=2$, $r_0=10$, $c=0.06$, $w_1=1$. In (a) we see the spectrum and the amplitude profile on the input plane. In (b) the three-dimensional wave dynamics are depicted along with the theoretical prediction for the trajectory (black-white dashed curve). In (c) the intensity contrast is presented as a function of the propagation distance along with the theoretical prediction (shown in circles). In the last two rows we depict the intensity profile of the horizontal cross sections shown in (b) with the theoretical predictions shown in circles. Specifically, in the third row the cross sections are taken before the focus and the analytic prediction is obtained from Eq.~(\ref{eq:aaf:drhodr}). In the last row (f) is computed exactly at the focus and (g) after the focus while the theoretical estimates are given by Eq.~(\ref{eq:aaf:asy:006}). \label{fig:aaf:1}}
\end{figure}
\begin{figure}[t]
\includegraphics[width=0.9\columnwidth]{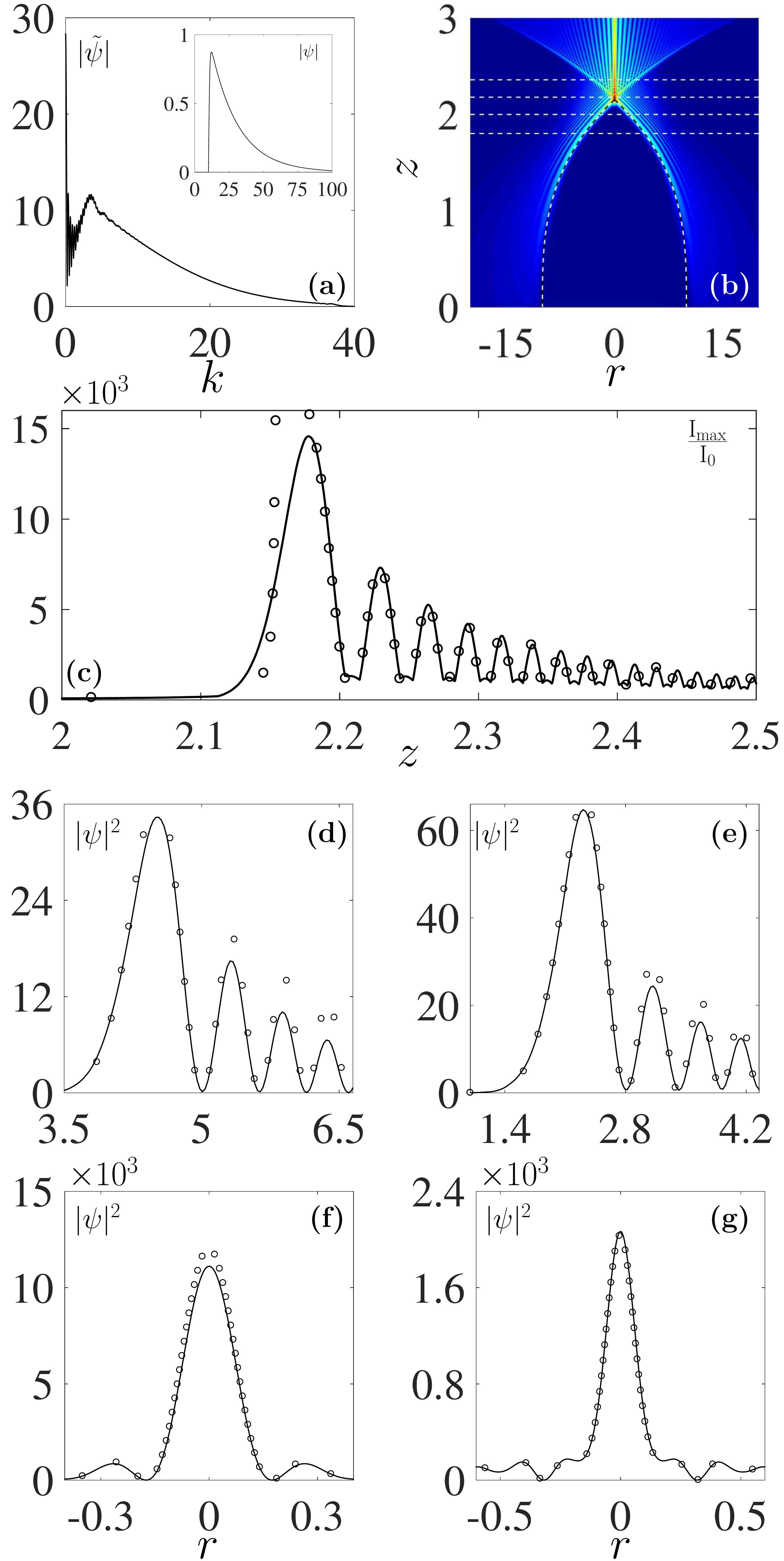} 
\caption{Same as in Fig.~\ref{fig:aaf:1} for a cubic trajectory ($\alpha=3$) with $\beta=1$, $r_0=10$, $c=0.05$, $w_1=1$.\label{fig:aaf:2}}
\end{figure}
In our simulations below we consider the case of abruptly autofocusing beams with power-law trajectories of the form
\begin{equation}
r_0-r_c=\beta z_c^\alpha
\label{eq:aaf:traj}
\end{equation}
where $r_0$ is the radius of the Airy ring on the input plane. 
The resulting phase is 
\begin{equation}
\phi(\rho) = 
\frac{-k\beta^{1/\alpha}\alpha^2}{(\alpha-1)^{1-1/\alpha}}
\frac{(\rho-r_0)^{2-1/\alpha}}{2\alpha-1}.
\label{eq:aaf:powerlawphase}
\end{equation}

In Fig.~\ref{fig:aaf:1} we see a typical example of an abruptly autofocusing wave with a parabolic trajectory and an exponential truncation of the form
\begin{equation}
A(r) = A_0
\sig\left(\frac{r-r_0}{w_1}\right)
\sig\left(\frac{r_a-r}{w_2}\right)
e^{c(r_0-r)},
\label{eq:aaf:ampl1} 
\end{equation}
where we define the sigmoid function $\sig$ as
\[
\operatorname{sig}(x) = 
\left\{
\begin{array}{cc}
\tanh(x) & x\ge0 \\
0 & x<0
\end{array}
\right.,
\]
$w_j$ are the slopes of the sigmoid functions, and $r_a$ is the selected aperture. 
In Fig~\ref{fig:aaf:1}(c) we compare the numerically derived maximum intensity contrast in the transverse plane as a function of the propagation distance with the theoretical prediction. Specifically, we utilize Eq.~(\ref{eq:aaf:global}) for $z\le  z_f$ and Eq.~(\ref{eq:aaf:asy:006}) for $z>z_f$. We see that our theoretical results are in good agreement with the numerical simulations. Some deviations appear in the slope of the maximum amplitude just before the focus. Specifically, the theoretical curve is steeper as compared to the numerical curve. In addition, the theoretical prediction gives a slightly higher intensity contrast at the focus. We attribute both of these differences to diffraction effects that are not taken into account in the theoretical calculations. In the third row of Fig.~\ref{fig:aaf:1} we see typical cross sections of the beam intensity before the focus. The numerical results are compared with the theoretical formula given by Eq.~(\ref{eq:aaf:drhodr}). The agreement is very good in capturing the behavior of the first Airy lobe whereas deviations in the amplitude start to appear in the subsequent lobes. In the forth row of Fig.~\ref{fig:aaf:2} we show the transverse beam amplitude at the focus (f) and after the focus (g). The theoretical results provided by Eq.~(\ref{eq:aaf:asy:006}) compare quite well with the numerical simulations. In Fig.~\ref{fig:aaf:2} we see similar results in the case of a cubic trajectory. The higher contrast is attributed to the increased value of the curvature at the focus and the larger value of $\rho_c$.

\begin{figure}
\centerline{\includegraphics[width=\columnwidth]{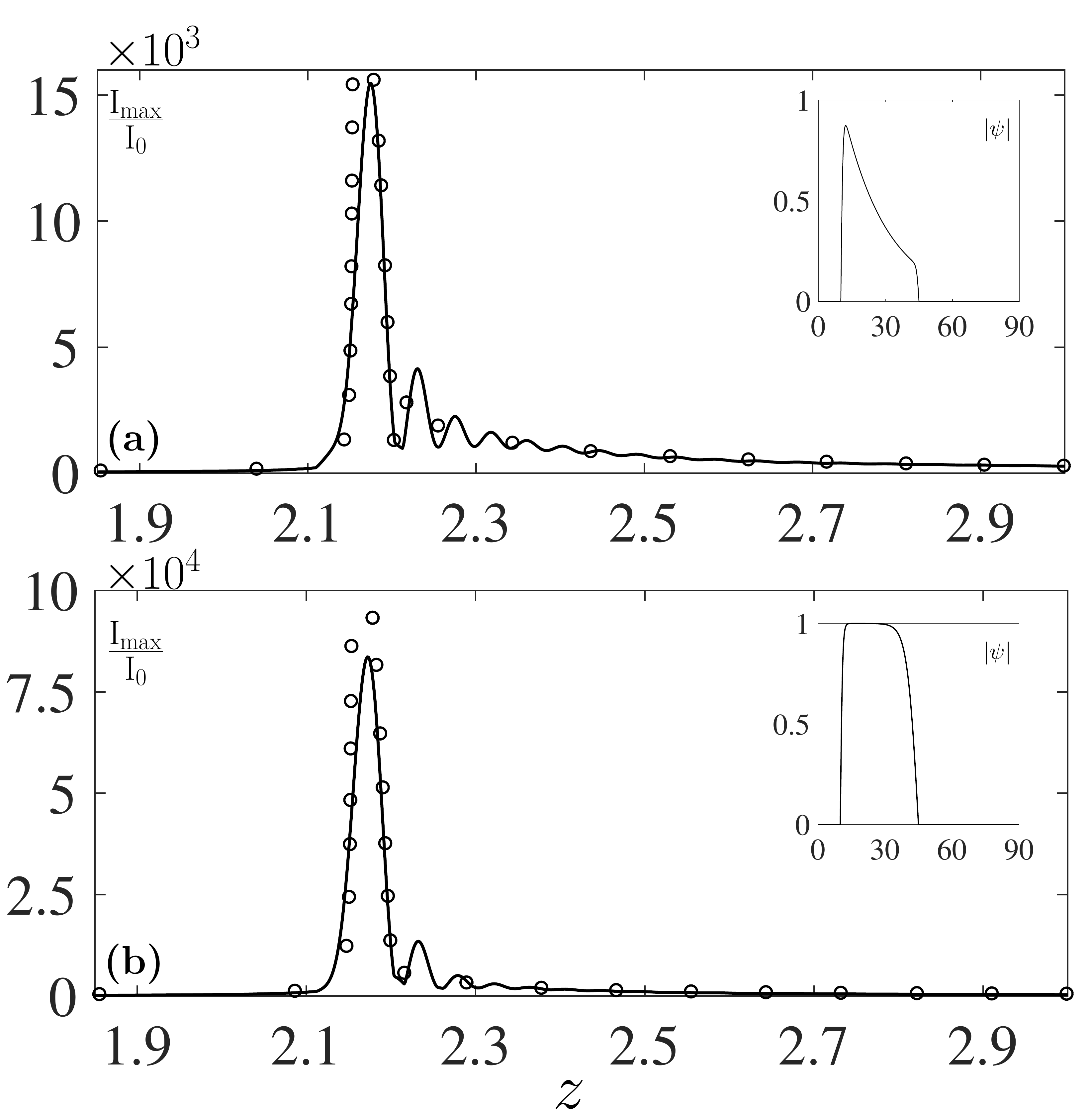}}
\caption{Intensity contrast as a function of the propagation distance. In the insets the initial conditions are shown. The parameters for the trajectory are the same as in Fig.~\ref{fig:aaf:2}. In (a) the amplitude on the input plane is the same as in Fig.~\ref{fig:aaf:2} but the aperture is reduced to $r_a=45$ and $w_2=1$. In (b) the amplitude is constant $c=0$ while $r_a=45$ and $w_2=5$. 
\label{fig:aaf:3}}
\end{figure}
We would like to optimize the properties of the abruptly autofocusing beams by (a) reducing the intensity of the oscillations that take place after the focus and (b) increasing the contrast at the focus. We take as a reference the results of Fig.~\ref{fig:aaf:2} and select to keep the same caustic trajectory. We can achieve (a) (reduced intensity after the focus) by decreasing the aperture as much as possible as long as it does not significantly affect the contrast (due to diffraction effects). In Fig.~\ref{fig:aaf:3} the intensity drops much faster after the focus due to the reduced value of the aperture. Considering point (b), from Eq.~(\ref{eq:foc_ampl}) we can compute the intensity contrast at the focus as
\begin{equation}
\frac{I_{\mathrm{max}}(z=z_f)}{I_{\mathrm{max}}(z=0)}
\approx
17.98
\left(
\frac{(k^2\kappa)^{1/3}\rho_c A(\rho_c)}{\max(A(\rho))}
\right)^2. 
\end{equation}
We see that the contrast depends on the curvature of the trajectory at the focus $\kappa(\rho_c)$, on $\rho_c$, and on the fraction $A(\rho_c)/\max(A(\rho))$. 
Both $\kappa(\rho_c)$ and $\rho_c$ depend on the geometric properties of the caustic trajectory -- a different trajectory with increased values of $\kappa$ and $\rho_c$ is going to exhibit increased focal contrast. However we can use the same trajectory and still achieve increased contrast by increasing the value of $A(\rho)/\max(A)$ up to unity. Specifically, in our simulation shown in Fig.~\ref{fig:aaf:3}(b) we select to keep a constant amplitude $A$ on the input plane in order to diminish possible diffraction effects. We clearly see a significant enhancement of the intensity contrast at the focus and a fast decrease in the intensity oscillations after the focus. 


\section{Implementation}

An important question is whether there are efficient methods to experimentally observe the families of optical waves discussed in this paper. In this respect, there are several works that have proposed methods to encode both amplitude and phase information by modulating only one of these two degrees of freedom. Such methods result in significant reduction of the experimental complexity. In~\cite{lee-ao1979,lee-po1978} different techniques are suggested that allow for the storage of both amplitude and phase information into binary computer generated holograms. For example, for an aperture function
\begin{multline*}
\frac{1}{2}\left\{
1+\sgn\left[
\cos\left(
\frac{2\pi x}{L}+\phi(x,y)
\right)-\cos\pi q(x,y)
\right]
\right\}=\\
\sum_{n=-\infty}^\infty
q(x,y)\operatorname{sinc}\left(\pi n q(x,y)\right)
\exp\left[i n\left(\frac{2\pi x}{L}+\phi(x,y)\right)\right],
\end{multline*}
where
\[
\frac1\pi\sin \pi q(x,y) = A(x,y),
\]
$\sinc(x)=\sin(x)/x$, and $\sgn$ is the sign function, we see that the first diffraction order reproduces both the amplitude and the phase of an optical wave. This technique has been utilized to generate different classes of nonparaxial accelerating plasmon beams~\cite{libst-prl2014}. 

In addition, both amplitude and phase information can be encoded into a phase only filter~\cite{davis-ao1999}. In particular, a phase pattern of the form $e^{iA(x,y)\phi(x,y)}$ with the phase been spatially modulated is used and the desired waveform is obtained in the first diffraction order. Such a configuration has been applied for the generation of abruptly autofocusing waves~\cite{papaz-ol2011}. 

\section{Conclusions}

In conclusion, we have shown that it is possible to generate beams with engineered trajectory/beam width and maximum amplitude along the trajectory. In addition in the case of abruptly autofocusing waves we are able to predict the amplitude profile along the trajectory and the intensity contrast but more importantly we are able to optimize the focusing procedure by revealing the particular parameters that should be taken into account. 
The results of the asymptotic calculations are expressed in an elegant form in terms of the parameters of the trajectory. 
Our results might be useful in areas where precise beam control is important such as particle manipulation and micromachining.

\section{Acknowledgments}

M.G. was supported by the Greek State Scholarships Foundation (IKY).
N.K.E. was supported by the Erasmus Mundus NANOPHI Project (2013-5659/002-001).


\newcommand{\noopsort[1]}{} \newcommand{\singleletter}[1]{#1}

\end{document}